\documentclass[conference]{IEEEtran}
\IEEEoverridecommandlockouts
\usepackage{cite}
\usepackage{amsmath,amssymb,amsfonts}
\usepackage{algorithmic}
\usepackage{graphicx}
\usepackage{textcomp}
\usepackage{xcolor}
\usepackage{ulem}
\usepackage{hyperref}
\def\BibTeX{{\rm B\kern-.05em{\sc i\kern-.025em b}\kern-.08em
    T\kern-.1667em\lower.7ex\hbox{E}\kern-.125emX}}
    
\begin{document}

% \title{Survey of Embedding Methods For Semantic Text Similarity In Bug Reports}

% \title{Enhancing Duplicate Bug Report Identification with Text Embedding Models: A Comparative Study}

\title{Comparative Analysis of Text Embedding Models for Bug Report Semantic Similarity}

\author{\IEEEauthorblockN{1\textsuperscript{st} Avinash Patil}
\IEEEauthorblockA{
\textit{Juniper Networks}\\
Sunnyvale, USA \\
patila@juniper.net}

\and
\IEEEauthorblockN{2\textsuperscript{nd} Kihwan Han}
\IEEEauthorblockA{\textit{Juniper Networks} \\
Sunnyvale, USA\\
kihwanh@juniper.net}

\and
\IEEEauthorblockN{3\textsuperscript{rd} Aryan Jadon}
\IEEEauthorblockA{\textit{Juniper Networks} \\
Sunnyvale, USA \\
aryanj@juniper.net}
}

\maketitle

\begin{abstract}

Bug reports are an essential aspect of software development, and it is crucial to identify and resolve them quickly to ensure the consistent functioning of software systems. Retrieving similar bug reports from an existing database can help reduce the time and effort required to resolve bugs. In this paper, we compared the effectiveness of semantic textual similarity methods for retrieving similar bug reports based on a similarity score. We explored several embedding models such as TF-IDF (Baseline), FastText, Gensim, BERT, and ADA. We used the Software Defects Data containing bug reports for various software projects to evaluate the performance of these models. Our experimental results showed that BERT generally outperformed the rest of the models regarding recall, followed by ADA, Gensim, FastText, and TFIDF. Our study provides insights into the effectiveness of different embedding methods for retrieving similar bug reports and highlights the impact of selecting the appropriate one for this task. Our code is available on \href{https://github.com/av9ash/DuplicateBugDetection}{GitHub}.

\end{abstract}


\section{Introduction}

Bug reports are a crucial communication between developers and users for reporting bugs and requesting their resolution. Retrieving similar bug reports from an existing database can help reduce the time and effort required to resolve bugs \cite{b14}. However, manually identifying similar bug reports can take time and effort. A popular approach to identifying similar bug reports is using semantic textual similarity (STS) methods that measure the similarity between two texts based on their semantic meaning. These methods use several natural language processing (NLP) techniques and machine learning models trained on large text data corpora.

This paper compares the effectiveness of various semantic textual similarity methods for retrieving similar bug reports based on a similarity score. For this purpose, we explore several neural network text embedding models such as ADA (GPT3.5), FastText, Gensim, and BERT. ADA (by Open AI) \cite{b20} is a large language model for text search, text similarity, and code search. FastText (by Facebook) \cite{b18} is a neural network-based model that has been specifically designed for text classification tasks. Gensim (by Radim Rehurek) \cite{b19} is a topic modeling library used for the semantic analysis of text data. BERT (by Google) \cite{b17} is a state-of-the-art deep learning model showing promising results in various NLP tasks.

We used the Software Defect Datasets \cite{b11} \cite{b13}, which contain bug reports for various software projects. We investigated the performance of the models and compared their recall scores. Our study provides insights into the effectiveness of different embedding methods for retrieving similar bug reports and highlights the importance of selecting an appropriate model for this task. The remainder of this paper is organized as follows: Section 2 discusses related work in this area. Section 3 describes the dataset and experimental setup. Section 4 presents the results of our experiments. Section 5 discusses enhancements to STS for bug reports. Finally, Section 6 provides conclusions. The code implementation of experiments is available at \href{https://github.com/av9ash/DuplicateBugDetection}{https://github.com/av9ash/DuplicateBugDetection}

\subsection{Terminology}

In the field of studies, a bug report (BR) can be alternatively known as a problem report or defect report. A duplicate or child bug report indicates that it has been recognized as a replica of a previously submitted bug report. The original bug report is the one that has been identified as the first report of a bug, and it is mapped to one or more replicas. A bug report can also be referred to as a master or parent report (PR). Multiple child replicas are siblings to each other, having a common parent. Child reports are mapped to respective parent reports using a hash map discussed later. Reports that have never been linked as child or parent to any other bug report are classified as unique reports.

% Tasks:
% Performance Analysis on various k (1,250)
% Performance Analysis of various algorithms for IR on all of the data.
% Performance Analysis of various algorithms for IR on chunks of the data.
% Performance Analysis of various algorithms for Recall, Map and Reduction Rate.
% Propose a standard-like method to evaluate the performance of these models.

\section{Previous Work}

Wang et al. \cite{b1} proposed a new approach to duplicate bug report detection that uses both bug information and execution information. Bug information includes the summary and description of the bug. Execution information includes the steps to reproduce the bug and the functions that are called during the execution. The bug information was used to build a vector of words, while the execution information was used to build a vector of functions. The two vectors were then compared to an existing pair using a similarity measure, such as cosine similarity \cite{b1}. 

In \cite{b2}, Sun et al. proposed two methods to improve the accuracy of duplicate bug retrieval. First, they extended BM25F, a textual similarity measure originally designed for short unstructured queries, to BM25Fext. BM25Fext was specially designed for lengthy structured report queries by considering the weight of terms in queries. Second, they proposed a new retrieval function named REP that utilized other information in reports, such as product, component, priority, and a few other details. They optimized REP based on a training set using a two-round gradient descent algorithm that contrasts similar pairs of reports against dissimilar ones. Overall, the BM25Fext improved recall rate@k by 3–13\% and Mean Average Precision (MAP) by 4–11\% over BM25F.

Jalbert and Weimer \cite{b3} proposed a system that automatically identifies and filters out duplicate bug reports. This system used a variety of features, including the surface features of the bug report (e.g., the title and description), the textual semantics of the bug report (e.g., the keywords used), and the graph structure of the bug report (e.g., the relationships between bug reports). The authors evaluated their system on a dataset of 29,000 bug reports from the Mozilla project. Their system identified and filtered out 8\% duplicate bug reports  \cite{b3}.

Sureka and Jalote \cite{b4} investigated text mining-based approaches to analyze bug databases to uncover exciting patterns. Their approach used character-level representation. The approach has two main benefits. First, it is not dependent on any particular language, and it does not need specific preprocessing. Second, it can identify sub-word features, which is useful when comparing noisy text. The approach was evaluated on a database containing over 200,000 bug reports from the open-source Eclipse project. The results showed that, for 1,100 randomly selected test cases, the recall @ 50 was 33.92\%. For 2,270 randomly selected test cases with a title-to-title similarity of more than 50, the recall rate was 61.94\% \cite{b4}.

DBR-CNN by Xie et al. \cite{b5} is a deep learning model to extract semantic representations of bug reports, and DBR-CNN enhanced the textual features with domain-specific information. The study compared DBR-CNN with other approaches and traditional CNN models, explored the impact of parameter settings, and validated the extensibility and flexibility of the approach with different word embeddings. DBR-CNN outperformed other approaches and traditional CNN models. The performance of DBR-CNN was affected by filter number, filter length, and the word embedding choice \cite{b5}.

Rakha, Bezemer, and Hassan \cite{b7} proposed a new evaluation method for the automated retrieval of duplicate issue reports, which used all available reports rather than a subset. They found that the traditional evaluation method overestimated performance by 17-42\%. The paper also showed that using the resolution field value of an issue report can significantly improve performance. The authors suggested that future studies report a range of values for performance metrics and use the proposed realistic evaluation method.

 Patil and Jadon \cite{b8}  proposed a method that considers both structured and unstructured information in a bug report, such as summary, description, severity, impacted products, platforms, and categories. It utilized a specialized data transformer, a deep neural network, and a machine learning approach that does not generalize to retrieve matching bug reports. Through various experiments with large datasets of thousands of bug reports, they demonstrated that the proposed method achieved an accurate retrieval rate of 70\% for recall@5.

To obtain the similarity of bug reports, Hu et al. \cite{b9} used four components: TF-IDF Vector, Word Embedding Vector, Bug Product \& Component, and Document Embedding Vector. Text information was extracted from bug reports and used to create bug documents. The final score was calculated by combining these four components, then recommended the most similar k bugs based on a given bug.

Building on the previous studies \cite{b1}-\cite{b9}, our work focuses on retrieving similar bug reports from a database using a similarity score. The uniqueness of our study is that we specifically compared the effectiveness of multiple embedding models, including TF-IDF, FastText, Gensim, BERT, and GPT3.5, while also exploring the optimization of the look-back period. The findings of our research provide valuable insights for selecting the most suitable method to achieve optimal results in retrieving similar bug reports from the database.

\section{Data and Experimental Setup}

This study used the Defects dataset \cite{b11} \cite{b13}. This dataset encompasses bug reports from multiple software projects: EclipsePlatform, MozillaCore, Firefox, JDT, and Thunderbird. The dataset comprised approximately 480,000 bug reports, each encompassing a summary, a description, and metadata attributes, including bug ID, priority, component, status, duplicate flag, resolution, version, created time, and time of resolution. Structured information, in addition to summary and description, helps improve accuracy \cite{b2}.

The training data comprised a collection of parent and unique bug reports for all experiments in this study. The test data set consists of child bug reports. Table \ref{table:ttdc} shows the count of bug reports used to train and test the models.

\begin{table}
\caption{Reports Count for Training \& Testing }
\centering
\begin{tabular}{|c|c|c|}
        \hline
        Dataset & Train Count & Test Count \\
        \hline
        Firefox &88374 &27440 \\
        Eclipse &71198 &13958\\
        MozillaCore &165773 &39296\\
        JDT &38184 &7112\\
        Thunderbird &23389 &9161\\
        \hline
        Total &386919 &96967\\
        \hline
\end{tabular}
\label{table:ttdc}
\end{table}

\subsection{Data Extraction}

We extracted the above data from the original train-test dataset, consisting of two columns: issue ID and duplicate. Each issue ID represents a unique bug report; not every bug report has a duplicate. However, for those bug reports with duplicates, one or many duplicates can be associated with them. The data was split into two files: train and test. Certain issue IDs can appear in both files, with the exact or different reported duplicates.

Additionally, some child bug reports may be mapped to a parent bug report that itself is a duplicate. The training data only should contain parent bug reports and unique bug reports, and the test data should only contain duplicate bug reports. To create such data, we built a duplicate-to-original hash map. For brevity, we will refer to this map as \textit{dup-org} map throughout the text.

To create the test dataset, we used the keys from the \textit{dup-org} map, resulting in a collection comprising solely of duplicate bug reports. Conversely, the remaining issue IDs, which are not present in the list of duplicates, were employed to construct the training data set. Therefore, it was crucial to have a robust \textit{dup-org} hash map to ensure the correct partitioning of the data into the appropriate sets. 

To construct the \textit{dup-org} map accurately, the following procedure was followed:
\begin{itemize}
\item Initially, duplicates and their respective parents were added from the original training set to an intermediate map called \textit{dup-org-train}. Before adding a duplicate, we checked if it already existed in the \textit{dup-org-train} map. If a duplicate was already present, it indicates two different parent issues reporting it as a child.
\item When a duplicate was already present in the \textit{dup-org-train} map, it signified that all but one (the one with the lowest value Issue ID) of these parents were duplicates, often referred to as siblings. In such cases, we added the sibling as a duplicate by identifying the natural parent from the previous entry.
\item We repeated the exact process to create another intermediate map called \textit{dup-org-test} from the original test set.
\item Finally, the two intermediate dictionaries, \textit{dup-org-train} and \textit{dup-org-test}, were merged using the same procedure. The procedure above ensured that siblings were properly identified and accounted for at each step.

Once we obtained the final \textit{dup-org} map, we utilized it to create the training and test datasets. Furthermore, it was employed to evaluate the accuracy of different models. This methodology ensured the accurate identification of duplicates and siblings while building the \textit{dup-org} map, enabling the creation of reliable training and test datasets and facilitating accurate model evaluation.
\end{itemize}

\subsection{Preprocessing}
% Add in the limitations of the study that changing these models will impact the results as we are essentially comparing the pre-trained models only.

We generated embeddings for the bug reports using multiple methods, such as TF-IDF \cite{b24}, BERT \cite{b17}, Fasttext \cite{b18}, Doc2Vec \cite{b19}, and ADA (GPT3.5) \cite{b20}. To establish a baseline model, we specifically utilized TF-IDF embeddings. In numerous research studies, TF-IDF is widely adopted as a text representation technique, making it a well-established and commonly used reference point. By incorporating Scikit-learn's TF-IDF as the baseline, we were able to effectively compare the performance of other models against this established standard. This comparative analysis enabled us to evaluate the effectiveness of alternative approaches by contrasting their performance with the TF-IDF baseline. TF-IDF is expressed as: 

% ADA (by Open AI) \cite{b20} is a large language model for text search, text similarity, and code search. FastText (by Facebook) \cite{b18} is a neural network-based model that has been specifically designed for text classification tasks. Gensim (by Radim Rehurek) \cite{b19} is a topic modeling library used for the semantic analysis of text data. BERT (by Google) \cite{b17} is a state-of-the-art deep learning model showing promising results in various NLP tasks.

\begin{equation} \mathrm{TF-IDF}=tf(t,d)*idf(t,D) \end{equation}

For pre-processing, we employed default techniques for TF-IDF, FastText, ADA, and BERT embeddings, which include lower-casing, tokenization, and stop-word removal. We used regular expressions to refine the tokenization process further, defining the token pattern and facilitating the extraction of English and alphanumeric words. On the other hand, BERT embeddings did not require any specific pre-processing steps. However, for Doc2Vec embeddings, the "simple\_preprocess" method provided by the Gensim library ensured effective data pre-processing.

\subsection{Embedding models}
GPT3.5, BERT, Fasttext, and Doc2Vec models required loading pre-trained models for their respective embeddings. Specifically, for BERT, we utilized the "all-mpnet-base-v2" model, optimized for various use cases and trained on a large and diverse dataset comprising over 1 billion training pairs. For ADA, we used "text-embedding-ada-002," a GPT3.5 large language model for text search, text similarity, and code search. For Fasttext, we employed the "crawl-300d-2M-subword" model, which consisted of 2 million word vectors trained with subword information on the Common Crawl dataset, encompassing 600 billion tokens. In the case of Doc2Vec, we used the "GoogleNews-vectors-negative300" model, trained on a portion of the Google News dataset containing approximately 100 billion words. This model provided 300-dimensional vectors for 3 million words and phrases.

ADA, BERT, and Fasttext models are utilized without fine-tuning, while the Gensim model is fine-tuned specifically for the training PRs for each bug repository allowing us to leverage the strengths of these pre-trained models in our analysis.

\subsection{Training \& Testing}
The training mechanism for the Information Retrieval (IR) \cite{b16} model remains consistent and straightforward throughout this study. We employed the non-generalizing Nearest Neighbors model, which operates based on the principle of identifying the specified number of training samples that were closest in distance to a new point, utilizing a distance metric. Smaller distances indicate a higher degree of similarity between the points. We fitted this model with the training data embeddings from all the considered encoders, ensuring that the model incorporated the encoded information for effective retrieval and matching.

During testing, we queried the trained model using test data embeddings to obtain the top "n" matches. A query is successful if the known parent report ID is among the returned recommendations.

\subsection{Experiments}
In this paper, we addressed the following research questions (RQs):
\begin{itemize} 
\item RQ1: We retrieved various top $n$ recommendations for duplicate parent reports (PRs) from a collection of all Parent and Unique PRs. We evaluated the accuracy of each model across multiple values of the number of recommendations made, denoted by n. $n \epsilon [1, 5, 10, 15, ..., 495, 500]$,  allowing us to observe how the model's performance evolved as we expanded the pool of potential matches, providing a clearer understanding of the model's accuracy and effectiveness.
\item RQ2: We compared the recall accuracy of all models across five bug repositories for the top 5 recommendations, known as recall@5. 

Recall rate can be defined as: 

\begin{equation} Recall(n) = \frac{{\sum\nolimits_{i = 1 \ldots \# duplicates} {listed} (n)}}{{\# actual\,duplicates}}\end{equation}

This experiment's training and testing data remained the same as in the previous experiment.
\item RQ3: We extracted the difference in days between the creation dates of Parent Bug Reports (BRs) and Child BRs. This analysis provided valuable insights into whether we should consider all existing BRs for a document search or limit the search area to a specific date range.
\item RQ4: We imposed a constraint on the search range by considering the creation dates of bug reports. Specifically, we limited the search for existing parent bug reports to include bug reports filed within the last ${d}$ days, as depicted in Table \ref{table:slder}. By reducing the number of bug reports to match with, we aimed to achieve better search results and minimize false positives.
\end{itemize}

\section{Results}
\subsection{RQ1: Evaluation of model performance for various incremental values of the recall rate}
\begin{figure}[htbp]
\centerline{\includegraphics[width=0.5\textwidth]{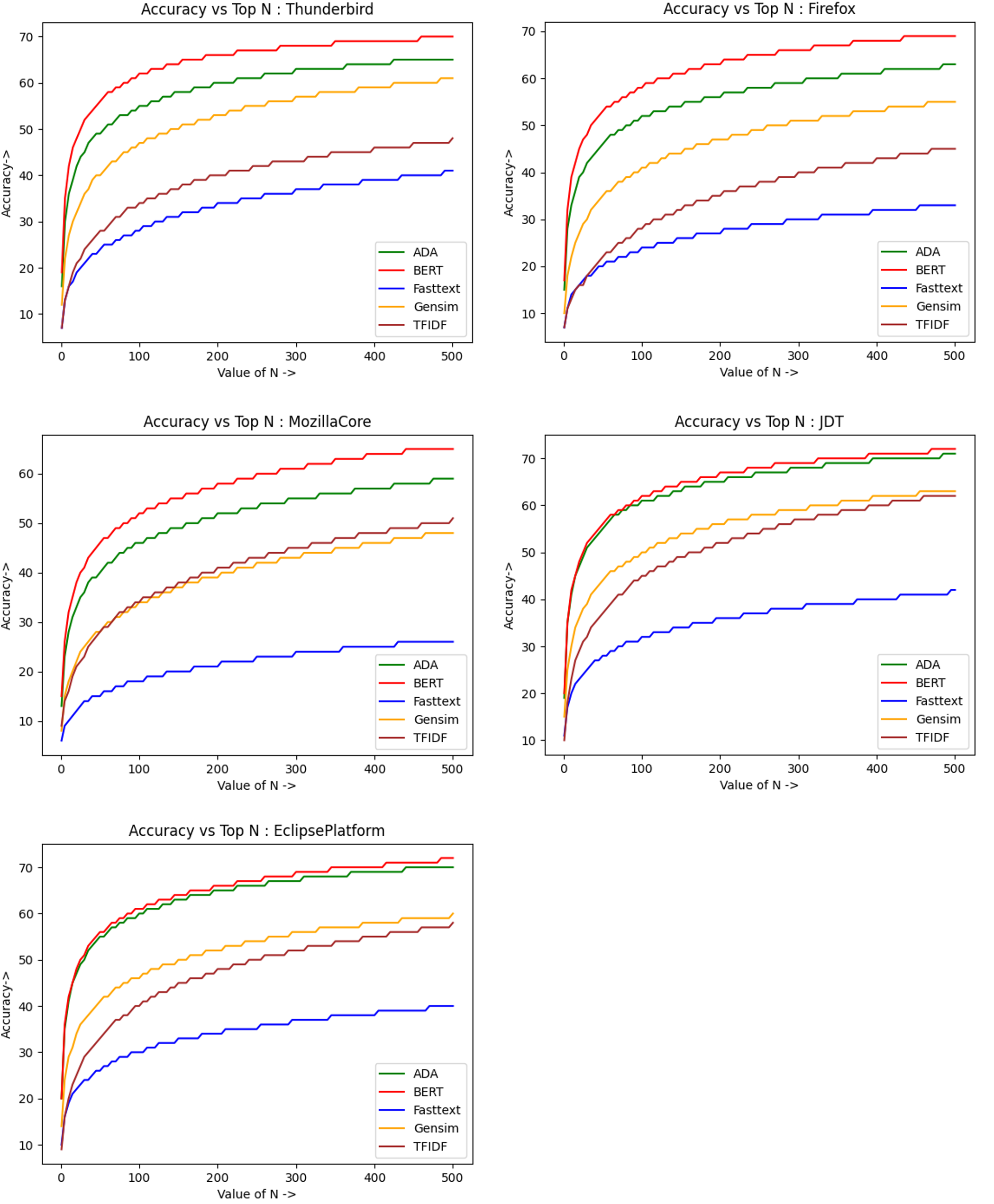}}
\caption{Accuracy vs Number of Recommendations.}
\label{fig:avnrall}
\end{figure}

Fig \ref{fig:avnrall} provides an initial insight into model comparison, revealing a clear performance order as follows: $BERT > ADA > Gensim > TFIDF > Fasttext$. Additionally, it is noticeable that the accuracy change was more prominent at smaller values of $n$ and tends to flatten out as $n$ increases, suggesting that fetching a more significant number of potential matches does not necessarily result in a significant increase in accuracy.

% \begin{figure}[htbp]
% \centerline{\includegraphics[width=0.5\textwidth]{Thunderbird_500.png}}
% \caption{Accuracy vs Number of Recommendations on Thunderbird.}
% \label{fig:avnr1}
% \end{figure}

% \begin{figure}[htbp]
% \centerline{\includegraphics[width=0.5\textwidth]{JDT_500.png}}
% \caption{Accuracy vs Number of Recommendations on JDT.}
% \label{fig:avnr2}
% \end{figure}

% \begin{figure}[htbp]
% \centerline{\includegraphics[width=0.5\textwidth]{EclipsePlatform_500.png}}
% \caption{Accuracy vs Number of Recommendations on EclipsePlatform.}
% \label{fig:avnr3}
% \end{figure}

% \begin{figure}[htbp]
% \centerline{\includegraphics[width=0.5\textwidth]{Firefox_500.png}}
% \caption{Accuracy vs Number of Recommendations on Firefox.}
% \label{fig:avnr4}
% \end{figure}

% \begin{figure}[htbp]
% \centerline{\includegraphics[width=0.5\textwidth]{MozillaCore_500.png}}
% \caption{Accuracy vs Number of Recommendations on MozillaCore.}
% \label{fig:avnr5}
% \end{figure}

% Experiment 2: Retrieving Top5 duplicate PRs from a collection all Parent PRs
\subsection{RQ2:  Comparison of recall accuracy of models across bug repositories for Top 5 Recommendation (recall@5)}
\begin{figure}[htbp]
\centerline{\includegraphics[width=0.5\textwidth]{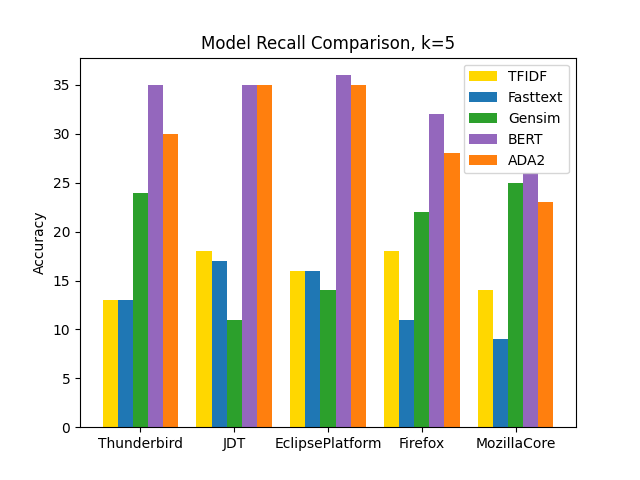}}
\caption{Recall Comparison for Each Bug Database.}
\label{fig:acmp}
\end{figure}

 Based on the findings presented in Fig \ref{fig:acmp}, it is evident that BERT consistently outperformed the other models regarding recall accuracy. On the other hand, Fasttext exhibited lower accuracy compared to the baseline TFIDF model. These results provide insights into the comparative performance of the models. 

% Experiment 3: Retrieving Top5 duplicate PRs from a collection all Parent PRs + Unique PRs
\subsection{RQ3: Analysis of Creation Date differences between Parent and Child Bug Reports and its impact on document search}

\begin{figure}[htbp]
\centerline{\includegraphics[width=0.5\textwidth]{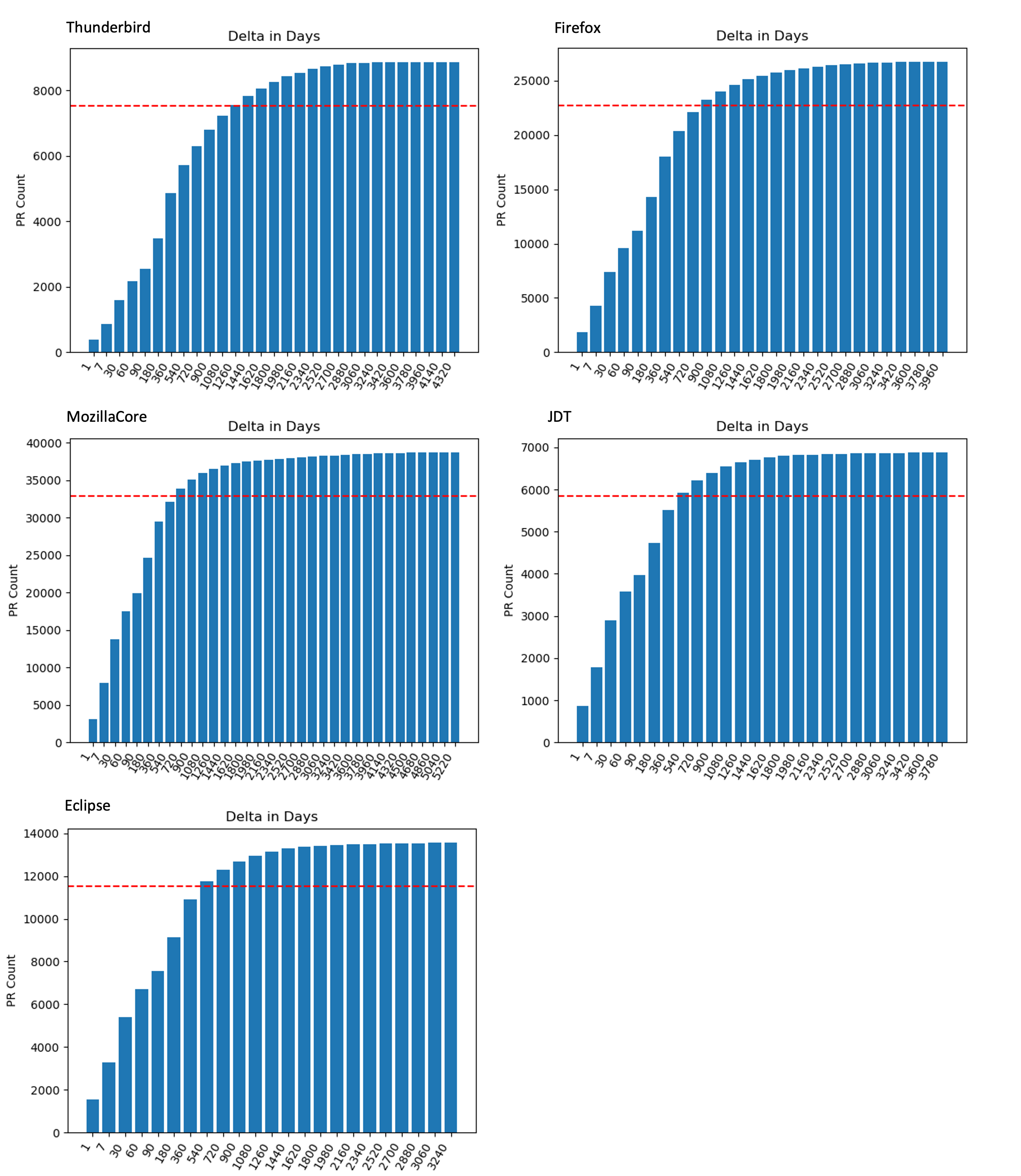}}
\caption{Duplicate PRs vs Creation Date Difference.}
\label{fig:didall}
\end{figure}

Fig \ref{fig:didall} assists in understanding the delta in creation dates, measured in days, between Parent and Child BRs. The Red dotted line represents 85\% of all duplicate BRs. Approximately 15\% of BRs in all projects exhibited significant time gaps between the first and second occurrences, ranging from 720 days to 5200 days. These substantial time gaps significantly increased the number of Bug Reports included in the document search. Alternatively, we can limit the search for Parent Bug Reports to a specific number of days in the past. This approach reduced the number of irrelevant BRs to consider in the search.

% \begin{figure}[htbp]
% \centerline{\includegraphics[width=0.5\textwidth]{Thunderbird_delta.png}}
% \caption{Duplicate PRs vs Creation Date Difference: Thunderbird}
% \label{fig:did1}
% \end{figure}

% \begin{figure}[htbp]
% \centerline{\includegraphics[width=0.5\textwidth]{JDT_delta.png}}
% \caption{Duplicate PRs vs Creation Date Difference: JDT.}
% \label{fig:did2}
% \end{figure}

% \begin{figure}[htbp]
% \centerline{\includegraphics[width=0.5\textwidth]{EclipsePlatform_delta.png}}
% \caption{Duplicate PRs vs Creation Date Difference: EclipsePlatform.}
% \label{fig:did3}
% \end{figure}

% \begin{figure}[htbp]
% \centerline{\includegraphics[width=0.5\textwidth]{Firefox_delta.png}}
% \caption{Duplicate PRs vs Creation Date Difference: Firefox.}
% \label{fig:did4}
% \end{figure}

% \begin{figure}[htbp]
% \centerline{\includegraphics[width=0.5\textwidth]{MozillaCore_delta.png}}
% \caption{Duplicate PRs vs Creation Date Difference: MozillaCore.}
% \label{fig:did5}
% \end{figure}

\subsection{RQ4:  Impact of Search Range Constraint on Bug Report Retrieval Accuracy Using Creation Dates.}
As shown in Table \ref{table:mrc} datasets marked with * represent the results obtained when the search was limited to the last ${d}$ days. This constraint improved the accuracy, as indicated by the higher number of instances where the accuracy has improved. These findings highlight the effectiveness of limiting the search range based on creation dates in enhancing the accuracy of the matching process. Nevertheless, it is crucial to consider the specific dataset and embedding model employed, as the results may vary depending on these factors.

\begin{table}
\caption{Search Limit in Days for Each Dataset}
\centering
\begin{tabular}{|c|c|}
        \hline
        Dataset & Search Limit (In Days) \\
        \hline
        Firefox &1080  \\
        Eclipse &720 \\
        MozillaCore &900 \\
        JDT &720 \\
        Thunderbird &1440\\
        \hline

\end{tabular}
\label{table:slder}
\end{table}

\begin{table}
\caption{Model Recall Comparison, k=5}
\centering
\begin{tabular}{|c|c|c|c|c|c|}
        \hline
        Dataset & TFIDF & Fasttext & Gensim & BERT & ADA \\
        \hline
        Firefox &13 &13 &24 &35 &28\\
        Firefox* &19 &14 &6 &32 &32\\
        Eclipse &18 &17 &11 &35 &35 \\
        Eclipse* &22 &19 &13 &34 &38\\
        MozillaCore &16 &16 &14 &36 &23\\
        MozillaCore* &18 &11 &7 &28 &27\\
        JDT &18 &11 &22 &32 &35\\
        JDT* &24 &20 &13 &34 &38\\
        Thunderbird &14 &9 &25 &26 &30\\
        Thunderbird* &15 &14 &6 &31 &31\\
        \hline
\end{tabular}
\vspace{1ex}
{\par *Search constrained by Creation Date as per Table II. \par}
\label{table:mrc}
\end{table}

\section{Discussion}
In this paper, we compared the effectiveness of  TF-IDF (Baseline), FastText, Gensim, BERT, and ADA as semantic textual similarity methods for retrieving similar bug reports. Our assessment revealed that BERT generally outperformed the rest of the models regarding precision and recall, followed by ADA, Gensim, FastText, and TFIDF.

Based on the assessments, we identify and propose several ways to efficiently evaluate Semantic Text Similarity (STS) in Bug Reports Databases. Building on top of the evaluation method proposed in \cite{b7}, we assert that the evaluation should focus on incorporating a significant number of Bug Reports in the search or training set, closely resembling the number of Bugs in the production instance of the Bug Tracking System. We recommend encompassing both Parent and Unique Bug Reports in the search or training set. Studies like \cite{b8}, \cite{b6}, and \cite{b10} that solely considered Parent Bug Reports in the search space failed to replicate real-world production scenarios, where any existing report could match a new/query bug report.

Restricting the search space to a small set of Parent and Unique Bug Reports did not account for the algorithm's potential confusion when numerous Bug Reports closely matched a new Bug report \cite{b4}, leading to a decrease in the ranking of the actual parent report Fig \ref{fig:avnrall}. Overall, this overestimated the performance of model \cite{b7}.

Compared to classification, measuring accuracy based on efficient retrieval is crucial, as done in \cite{b3}  because triage engineers have new bug reports to compare with all the existing ones \cite{b15}. Their task rarely involves bug report pairs that must be labeled as duplicates or not.

Additionally, searching through all existing Bug Reports may only sometimes yield optimal results, prompting the exploration of limiting the number of Bug Reports to be searched based on age. Older Bug Reports are less likely to encounter new duplicate reports. Moreover, it is vital to consider the practical aspects of the evaluation process. Training engineers and developers are less likely to review all 20 or 25 recommendations a duplicate detection algorithm provides. Therefore, the recall rate can be limited to only the top 5 recommendations.

Limitations of the study include that the effectiveness heavily depends on the choice of embedding models. Embedding models like BERT and ADA (GPT-3.5) might have evolved since the study's completion, potentially affecting their performance characteristics. Some models utilized in this study (ADA, BERT, and FastText) have been trained on diverse datasets, or have undergone fine-tuning (Gensim), potentially affecting their general performance and comparability. While we compare multiple models, there might be other advanced models that could yield different results.

\section{Conclusion}

In this study, we compared different models for retrieving top $n$ recommendations for duplicate PRs from all Parent and Unique PRs, observing changes in their performance by increasing the size of $n$. We compared recall accuracy across five bug repositories for the top 5 recommendations. Our results showed that BERT consistently outperformed other models, while Fasttext showed lower accuracy than the TFIDF baseline. In addition, we conducted an analysis to extract the difference in creation dates between Parent and Child BRs. The above analysis revealed that limiting the search to a specific range can reduce the number of irrelevant BRs and enhance search results.

We also proposed enhancements to evaluate the efficiency of STS in Bug Reports databases. We emphasized the importance of incorporating significant Bug Reports in the search or training set, including both Parent and Unique Bug Reports. Restricting the search space to a small set of Parent and Unique Bug Reports may overestimate model performance and fail to replicate real-world scenarios. Efficient retrieval and limiting Bug Reports based on their age were suggested for practical considerations, aiding triage engineers' tasks.

Overall, our study provides valuable insights into efficient evaluation methods for STS in Bug Reports databases, guiding the selection of appropriate models and evaluation techniques for better performance in real-world applications.


\begin{thebibliography}{00}

\bibitem{b1} X. Wang, L. Zhang, T. Xie, J. Anvik and J. Sun, "An approach to detecting duplicate bug reports using natural language and execution information," 2008 ACM/IEEE 30th International Conference on Software Engineering, Leipzig, Germany, 2008, pp. 461-470, doi: 10.1145/1368088.1368151.

\bibitem{b2} C. Sun, D. Lo, S. -C. Khoo and J. Jiang, "Towards more accurate retrieval of duplicate bug reports," 2011 26th IEEE/ACM International Conference on Automated Software Engineering (ASE 2011), Lawrence, KS, USA, 2011, pp. 253-262, doi: 10.1109/ASE.2011.6100061.

\bibitem{b3} N. Jalbert and W. Weimer, "Automated duplicate detection for bug tracking systems," 2008 IEEE International Conference on Dependable Systems and Networks With FTCS and DCC (DSN), Anchorage, AK, USA, 2008, pp. 52-61, doi: 10.1109/DSN.2008.4630070.

\bibitem{b4} A. Sureka and P. Jalote, "Detecting Duplicate Bug Report Using Character N-Gram-Based Features," 2010 Asia Pacific Software Engineering Conference, Sydney, NSW, Australia, 2010, pp. 366-374, doi: 10.1109/APSEC.2010.49.

\bibitem{b5} Q. Xie, Z. Wen, J. Zhu, C. Gao and Z. Zheng, "Detecting Duplicate Bug Reports with Convolutional Neural Networks," 2018 25th Asia-Pacific Software Engineering Conference (APSEC), Nara, Japan, 2018, pp. 416-425, doi: 10.1109/APSEC.2018.00056.

\bibitem{b6} P. Runeson, M. Alexandersson and O. Nyholm, "Detection of Duplicate Defect Reports Using Natural Language Processing," 29th International Conference on Software Engineering (ICSE'07), Minneapolis, MN, USA, 2007, pp. 499-510, doi: 10.1109/ICSE.2007.32.

\bibitem{b7} M. S. Rakha, C. -P. Bezemer and A. E. Hassan, "Revisiting the Performance Evaluation of Automated Approaches for the Retrieval of Duplicate Issue Reports," in IEEE Transactions on Software Engineering, vol. 44, no. 12, pp. 1245-1268, 1 Dec. 2018, doi: 10.1109/TSE.2017.2755005.

\bibitem{b8} A. Patil and A. Jadon, "Auto-labelling of Bug Report using Natural Language Processing," 2023 IEEE 8th International Conference for Convergence in Technology (I2CT), Lonavla, India, 2023, pp. 1-7, doi: 10.1109/I2CT57861.2023.10126470.

\bibitem{b9} D. Hu et al., "Recommending Similar Bug Reports: A Novel Approach Using Document Embedding Model," 2018 25th Asia-Pacific Software Engineering Conference (APSEC), Nara, Japan, 2018, pp. 725-726, doi: 10.1109/APSEC.2018.00108.

\bibitem{b10} N. Kaushik and L. Tahvildari, "A Comparative Study of the Performance of IR Models on Duplicate Bug Detection," 2012 16th European Conference on Software Maintenance and Reengineering, Szeged, Hungary, 2012, pp. 159-168, doi: 10.1109/CSMR.2012.78.

\bibitem{b11} A Jadon, A Patil, and S Jadon. "A Comprehensive Survey of Regression Based Loss Functions for Time Series Forecasting." 2022 arXiv preprint arXiv:2211.02989.

\bibitem{b12} S. Jadon and A. Jadon, "MetaForecast: Harnessing Model-Agnostic Meta-Learning Approach to Predict Key Metrics of Interconnected Network Topologies," 2023 IEEE International Conference on Industry 4.0, Artificial Intelligence, and Communications Technology (IAICT), BALI, Indonesia, 2023, pp. 239-243, doi: 10.1109/IAICT59002.2023.10205730.

\bibitem{b13} S. Jadon and A. Jadon, “An overview of deep learning architectures in few-shot learning domain,” 2023.

\bibitem{b14} N. Bettenburg, R. Premraj, T. Zimmermann and 3. Sunghun Kim, "Duplicate bug reports considered harmful … really?," 2008 IEEE International Conference on Software Maintenance, Beijing, China, 2008, pp. 337-345, doi: 10.1109/ICSM.2008.4658082.

\bibitem{b15} J. Anvik, L. Hiew and G. C. Murphy, "Who should fix this bug?", Proc. 28th Int. Conf. Softw. Eng., pp. 361-370, 2006.

\bibitem{b16} R. Baeza-Yates and W. B. Frakes, Information Retrieval: Data Structures \& Algorithms, Englewood Cliffs, NJ, USA:Prentice Hall, 1992.

\bibitem{b17} J Devlin, M W Chang, K Lee et al., BERT: Pre-training of Deep Bidirectional Transformers for Language Understanding[J], 2018.

\bibitem{b18} P. Bojanowski, E. Grave, A. Joulin, and T. Mikolov, "Enriching Word Vectors with Subword Information," Transactions of the Association for Computational Linguistics, vol. 5, pp. 135-146, 2017.

\bibitem{b19} R. Řehůřek and P. Sojka, "Software Framework for Topic Modelling with Large Corpora," in Proceedings of the LREC 2010 Workshop on New Challenges for NLP Frameworks, Valletta, Malta, May 22, 2010, pp. 45-50.

\bibitem{b20} A. Jadon and S. Kumar, "Leveraging Generative AI Models for Synthetic Data Generation in Healthcare: Balancing Research and Privacy," 2023 International Conference on Smart Applications, Communications and Networking (SmartNets), Istanbul, Turkiye, 2023, pp. 1-4, doi: 10.1109/SmartNets58706.2023.10215825.

\bibitem{b21} P.Avinash, "Bugrepo," [Online]. Available: https://github.com/av9ash/bugrepo. Accessed on: May 7, 2023.

\bibitem{b22} P.Avinash, "DuplicateBugDetection," [Online]. Available: https://github.com/av9ash/DuplicateBugDetection. Accessed on: May 7, 2023.89

\bibitem{b23} “BugRepo,” GitHub, Jul. 19, 2023. https://github.com/logpai/bughub (accessed Jul. 24, 2023).

\bibitem{b24} T. Brown et al., "Language models are few-shot learners," in Advances in Neural Information Processing Systems, vol. 33, pp. 1877-1901, 2020.

\bibitem{b25} F. Pedregosa, G. Varoquaux, A. Gramfort, V. Michel, B. Thirion, O. Grisel, M. Blondel, P. Prettenhofer, R. Weiss, V. Dubourg, J. Vanderplas, A. Passos, D. Cournapeau, M. Brucher, M. Perrot, and E. Duchesnay, "Scikit-learn: Machine Learning in Python," Journal of Machine Learning Research, vol. 12, pp. 2825-2830, 2011.

\end{thebibliography}
\end{document}